\documentclass[twoside]{article} 
\usepackage{fleqn,espcrc2} 
\usepackage{epsfig}

\newcommand{\fnm}[1]{\footnotemark[#1]}

\newcommand{\oomega}{\overline{\omega}_{\rm pl}}

\renewcommand{\u}[1]{ {\bf   #1   } }

\newcommand{\ukp}{\u{k}^{\prime}} 
 
\newcommand{\uk}{\u{k}} 
\newcommand{\cpd}[1]{ c^{\phantom \dagger}_{ #1}}

\newcommand{\tp}{ t^{\prime}}

\begin{document}


\title{Static Charge Coupling of Intrinsic Josephson Junctions} 

\author{Ch. Helm\address{Los Alamos National Laboratory, Division  
T-11, M.S. B-262, NM 87545, USA}, J. Keller\address{Institute of  
Theoretical Physics, University of 
Regensburg, D-93040 Regensburg, Germany},\renewcommand{\thefootnote}{\alph{footnote}}  Ch. Preis\fnm{2},  
A. Sergeev\address{Dept. ECE, Wayne State University, Detroit, MI 48202, USA}} 

\maketitle
 
\begin{abstract} 
A microscopic theory for the coupling of intrinsic Josephson oscillations  
due to charge fluctuations on the quasi two-dimensional superconducting  
layers is presented.  
Thereby in close analogy to the normal state the effect of the  
scalar potential on the transport current is taken into account 
consistently.  The dispersion of collective modes is derived and an  
estimate of the coupling constant is given.  
It is shown that the correct treatment of the quasiparticle current is   
essential  
in order to get the correct position of Shapiro steps. In this case the  
influence of the coupling on dc-properties like the  
$I-V$-curve  is negligible.  
\end{abstract}

\bigskip  
Pacs: 74.72.-h,74.80.Fp,74.50.+r,74.40.+k \newline  
Keywords: layered superconductors, intrinsic Josephson effect,  
SN-junction, Shapiro Steps

\begin{center}  
(Proceedings of HTS Plasma 2000 Symposion, Sendai August 22-24 2000, to  
appear in Physica C) 
\end{center}

\bigskip  


\section{INTRODUCTION}

Since the discovery of the intrinsic Josephson effect not only 
the typical properties of conventional junctions 
 were demonstrated \cite{kleiner1}, but  also 
unique and surprising features like the coupling of Josephson oscillations 
to phonons \cite{wir1} have been discovered.   
There has also been a considerable interest in  the influence of  
nonequilibrium effects on the  $I-V$-characteristic and  
collective modes  
\cite{tachiki1,artemenko_jos,artemenko_d,artemenko_physicaC,diss,preis_sandiego,ryndyk_jetp}, as the quasi two-dimensional superconducting layers are  
expected to be more sensitive to external perturbations than bulk materials. Recent measurements 
 of Shapirosteps on the resistive branch \cite{rother,doh2} 
suggest to study the role of non-equilibrium effects on these, as 
the in depth theoretical understanding of this problem is important for 
any high-precision applications of high temperature superconductors (HTSC)  
as a voltage standard.  
 
This paper is organized as follows:  
The microscopic theory for the 
electronic transport between the superconducting layers is developed and  
the close analogy of the static case to the normal state is pointed out.  
Then the   
consequences for the $I$-$V$-curve, the  
dispersion of collective modes and the position of Shapiro steps are being  
discussed.  
Further technical details can be found in \cite{diss,preis_sandiego} and in  
a forthcoming publication \cite{future}.


\section{TUNNELING THEORY}

\begin{figure}[htp!] 
\begin{center} 
\leavevmode
\epsfig{file=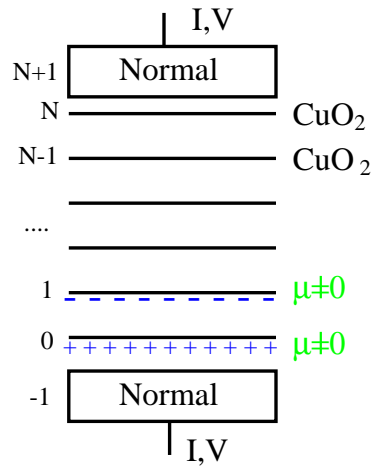,width=0.3\textwidth}\end{center} 
\caption{ \label{stackbild} 
Schematic picture of the mesa of superconducting layers in a 2-point  
measurement.  
 } 
\end{figure}

We consider a stack of $N+1$ (superconducting) layers  
$l=0,1,\dots,N$ forming $N$ intrinsic (Josephson) junctions in the  
homogeneous case.  The normal conducting  
electrodes attached at the top and bottom of the stack in a 2-point  
measurement are denoted as $l=-1,N+1$ (cf. Fig. \ref{stackbild}).  
 
As a motivation for the following discussion let us first recall  
the situation in  the normal state.  
There the electrical current (density)   
\begin{eqnarray} 
j_{l,l+1}^{NN} &=& {\sigma_{l,l+1}\over ed}  (\mu_{l+1}^{\rm ec}  
- \mu_l^{\rm ec} ) \\  
&=& \sigma_{l,l+1} E_{l,l+1} +  {\sigma_{l,l+1}\over ed}  
(\mu_{l+1}^c - \mu_l^c)   
\label{currentnormal2}  
\end{eqnarray}  
between the layers $l$ and $l+1$ is given  by the difference of  
the electrochemical potentials $\mu_l^{ec} = \mu^c_l - e A_{0,l} $ 
(Fermi energy) in neighbouring layers ($d$ distance of layers, 
$e=|e|$). This can be separated in a diffusion term  
driven by the difference $\Delta^{(1)} \mu^c_l := \mu_{l+1}^c - \mu_l^c $   
of the chemical potentials $\mu_l^c$ and a field term $\sigma_{l,l+1}  
E_{l,l+1} = \sigma_{l,l+1} ( A_{0,l} - A_{0,l+1}  ) $.  
 
In turn, the (static) charge fluctuation  
\begin{equation} 
\rho_{l,N} =  - 2  e  N_{2d} \mu_l^c  
\label{charge_normal}  
\end{equation}  
on the layer $l$ is determined completely by the filling of the conduction  
band or the chemical potential $\mu_l^c$ respectively. Using this  
 and the Poisson equation 
($\epsilon_{\infty}$: background dielectric constant) 
\begin{equation} 
    \rho_l = \epsilon_0 \epsilon_{\infty}  
\left( E_{l,l+1} - E_{l-1,l}  \right)  
\label{poisson}  
\end{equation}  
$\mu_l^c$ can be eliminated from equ. \ref{currentnormal2}: 
\begin{equation} 
j =  j_{l,l+1}^{NN} = \sigma_{l,l+1} \left( 1 - \alpha_0 \Delta^{(2)} \right) 
E_{l,l+1}   
\label{normalcurrent}  
\end{equation}   
with the discrete derivative $\Delta^{(2)} f_l :=  
f_{l+1} + f_{l-1} - 2 f_l $ and the coupling constant  
\begin{equation}  
\label{alpha_normal}
    \alpha_0 = \frac{\epsilon_0 \epsilon}{2 e^2 d N_{2d}} \ll 1 \quad .  
\end{equation} 
For a fixed dc-bias current $j$ equ. \ref{normalcurrent} 
can be used to determine the  
electric field $E_{l,l+1}$ by applying the operator $1 + \alpha_0  
\Delta^{(2)}$ on equ. \ref{normalcurrent}. If all conductivities  
$\sigma_{l,l+1}$ are equal,  no charge fluctuations accumulate and  
$E_{l,l+1} = j / \sigma  $, while in the case of only one barrier with a 
higher 
 resistance (e.g. $\sigma_{0,1} \ll \sigma_{l,l+1}, l \neq 0$ )
the electric field is not only localized at the highly resistive 
junctions, but is spread to neighboring junctions  
\begin{eqnarray} 
\label{E_normal1}
E_{0,1} &=& (1- 2 \alpha_0  ) j/ \sigma_{0,1} \quad ,  \\  
E_{-1, 0} &=& E_{1,2} = \alpha_0 j / \sigma_{0,1} \quad . 
 \label{E_normal2}
\end{eqnarray}

 
In the superconducting state both the electric field $E_{l,l+1}$ and the 
 difference ($l=0,\dots,N-1$) 
\begin{equation} 
\gamma_{l,l+1} = \chi_l - \chi_{l+1} - \frac{2e}{\hbar}  
\int_{l}^{l+1} A_z (z, t) dz    
\end{equation} 
of the phases $\chi_l$ of the superconducting order parameters  
$\Delta_l = | \Delta_l | e^{i\chi_l} $ in neighboring  
layers are gauge invariant quantities and have an independent physical  
meaning.  
Its time derivative 
directly leads to the general Josephson relation  
\begin{equation} 
\label{josephson} 
\frac{\hbar}{2e} {\dot \gamma}_{l,l+1} (t)  = V_{l,l+1} +  
\mu_{l+1} - \mu_l   
\quad  l=0,.,N-1    
\end{equation} 
between the phase $\gamma_{l,l+1}$ and the  
voltage $V_{l,l+1} = d_{l,l+1} E_{l,l+1}   $ across the junction, 
where the scalar potential $\mu_l$ is given by:  
\begin{equation} 
\mu_l = A_{0,l} - \frac{\hbar}{2e} {\dot \chi}_l (t) \quad \quad  
 l=0,1,\dots,N 
 \quad .  
\end{equation} 
As can be seen from the Gorkov equation for the single particle Greens function, 
the static part $\mu_l^{\rm dc} /e$ plays the role of the chemical potential 
$\mu_l^c$ in the superconducting state. 
 

In addition to this,  a tunneling Hamiltonian   
\begin{equation} 
\label{hamilton_tunnel} 
H_T = \sum_{l \u{r} \sigma} t_{ \u{r} }^{\phantom\ast}  
   c^{\dagger}_{l+1 \u{r} \sigma}  \cpd{l  \u{r} \sigma} 
   e^{ - i \frac{e}{\hbar} \int_{l}^{l+1}  
     A_z (z, t) \, dz } + c.c.  
\end{equation}   
 between the (BCS-like) superconducting layers has to be specified  
explicitly in order to calculate $j_{l,l+1}$.   
In the following we will assume a hopping matrix element $t_{\uk \ukp}^2 = 
t_s^2 + t_d^2 g_{\uk} g_{\ukp} $, which has both a constant $s$ and $d$-wave 
(i.e. $g_{\uk}$) part in order to get both a Josephson current and a finite 
quasiparticle conductivity at the same time.  
The general structure of the theory will be independent of this choice.

A time dependent  perturbation theory up to the second order in the  
hopping matrixelement $t_{\uk \ukp}$ is performed, but no assumption  
for the external electromagnetic potentials $A_{0.l} (t)$ and $A_z (t)$  
is made.  
The resulting expressions for the tunnel current  
\begin{equation}  
\label{current_general}
j_{l,l+1} = {\Im} \sum_{\pm}   \int\limits_{\infty}^t d t^{\prime}  
   P_{\pm}  ( t, \tp )  
 e^{\frac{i}{2} (  \gamma_{l,l+1} (t)  \pm  \gamma_{l,l+1} (\tp) ) } 
\end{equation} 
between the layers $l$ and $l+1$ and the charge fluctuation  
\begin{eqnarray} 
\rho_l &=& - \chi_{\rho \rho} \mu_l + Q_{t^2} (t, \gamma_{l,l+1} )  
\end{eqnarray}  
on layer $l$ are nonlinear in $\gamma_{l,l+1}$. 
Thereby  
$\chi_{\rho \rho}$ is the charge suszeptibility of the 
superconducting layer, which is for $\omega=0$ connected  with the two  
dimensional density of states $N_{\rm 2d}$ and independent of temperature  
$T$: $\chi_{\rho \rho} ( \omega = 0, T) = 2 e^2 N_{\rm 2d} $. 
The correlation functions $P_{\pm}$ and the charge contribution  $Q_{t^2}$  
in $O(t_{\uk \ukp}^2)$ also depend on the scalar potential $\mu_l$ and have 
been expressed by Feynman diagrams  in \cite{diss,preis_sandiego}.

Using the general ansatz for the phase  
\begin{equation} 
\gamma_{l,l+1} = \gamma_{0l} + \Omega_l t + \delta  
\gamma_{l,l+1} (t)  
\end{equation}  
in resistive ($\Omega_l \neq 0$) and superconducting ($\Omega_l 
 \neq 0$) junctions, $j_{l,l+1}$ and $\rho_l$ can  
be linearized in the (small) oscillations $\delta \gamma_{l,l+1}$ and  
the  scalar potential $\mu_l$:  
\begin{eqnarray} 
  j_{l,l+1}  &=& j_c \sin \gamma_{l,l+1} +  \frac{\hbar \sigma_0 }{2ed}  
    \dot{\gamma}_{l,l+1} -  
    \sigma_1 \frac{1}{d} \Delta^{(1)} \mu_l  
  \nonumber 
\\ 
&=& j_{\l,l+1}^{\rm sup} \! + \! \sigma_0 E_{l,l+1}  \!
 + \frac{( \sigma_0 - \sigma_1)}{d}  \Delta^{(1)} \mu_l  ,
 \label{strom_allgemein} 
 \\ \label{ladung_allgemein} \nonumber 
  \rho_{l}   &=& -  \chi_{\rho \rho} \mu_l + 
 \frac{i \sigma_2}{d \omega} \,  
  \Delta^{(2)} \mu_l - \frac{\hbar \sigma_1}{2ed}  \, 
   \Delta^{(1)} \gamma_{l-1,l}   \\  
&=& -  \chi_{\rho \rho} \mu_l  + \frac{i}{\omega } \left( j_{l,l+1}^Q -  j_{l-1,l}^Q
    \right) \label{chargegeneral} \quad , 
 \\ 
   j^{Q}_{l,l+1} &=&  \sigma_1 E_{l,l+1} +  
 \frac{ \left( \sigma_1 - \sigma_2  \right)}{ d}  
 \left(  \mu_{l+1} - \mu_l \right) \; . 
\end{eqnarray} 
Here we restricted the discussion to a linear quasiparticle characteristic 
for simplicity. The conductivities $\sigma_i ( \Omega_l , \omega )$ will in 
general depend on both $ \hbar \Omega_l /2e = 
V_{\rm dc} + \mu_{l+1}^{\rm dc} - 
\mu_l^{\rm dc}$ and the oscillation frequency $\omega$ of $\delta
 \gamma_{l,l+1}$, 
which coincides with $\Omega$ e.g. on the autonomous first resistive 
branch. Products like $( \sigma_0  E_{l,l+1} ) (t) = \int 
 \sigma_0 (t-t^\prime) E_{l,l+1} (t^\prime) d t^\prime$ are to be  
interpreted as a folding in time space.

\begin{figure}[htp!] 
\begin{center} 
\leavevmode 
\epsfig{file=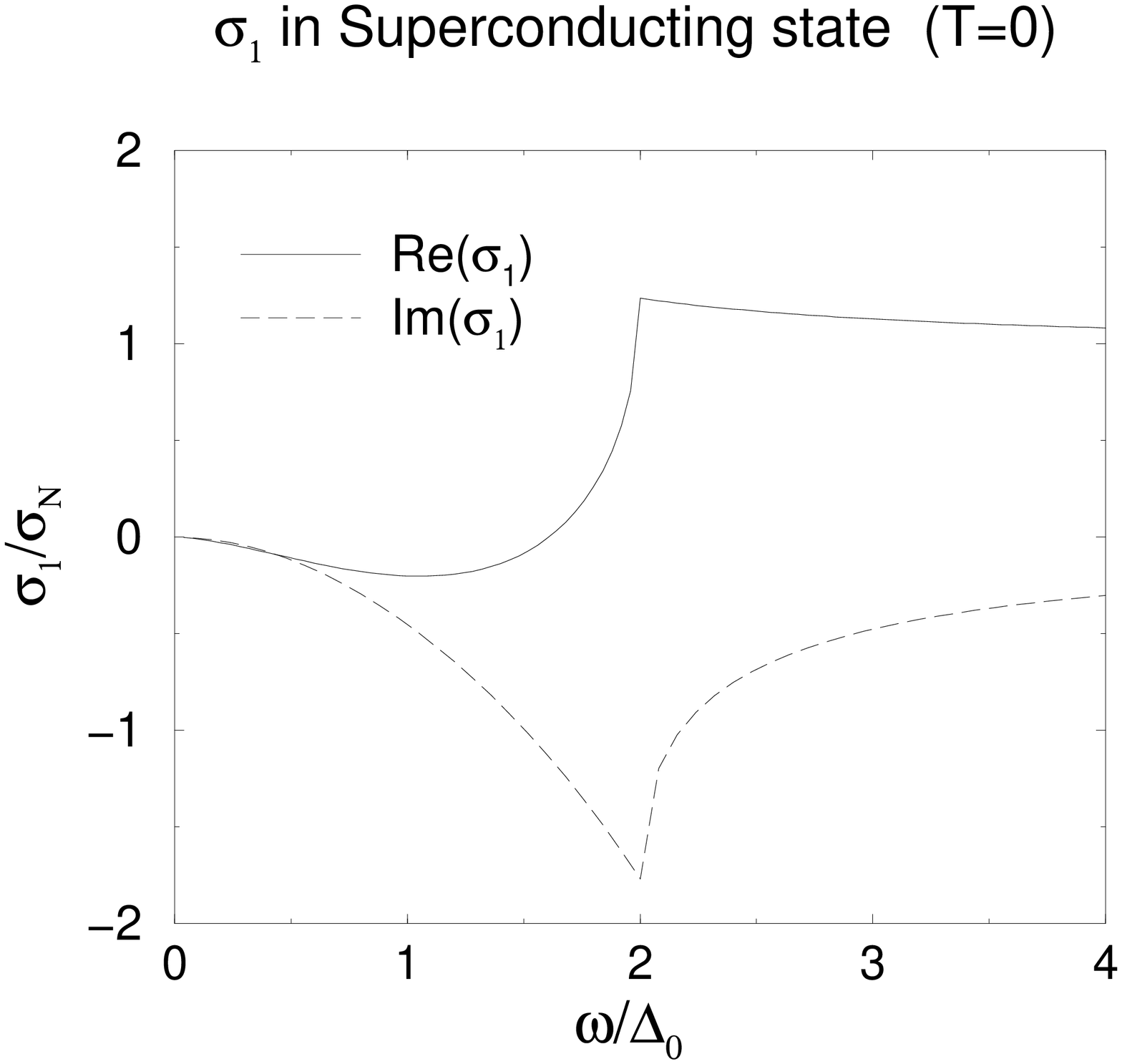,width=0.3\textwidth,angle=270,clip=} 
\newline  
\epsfig{file=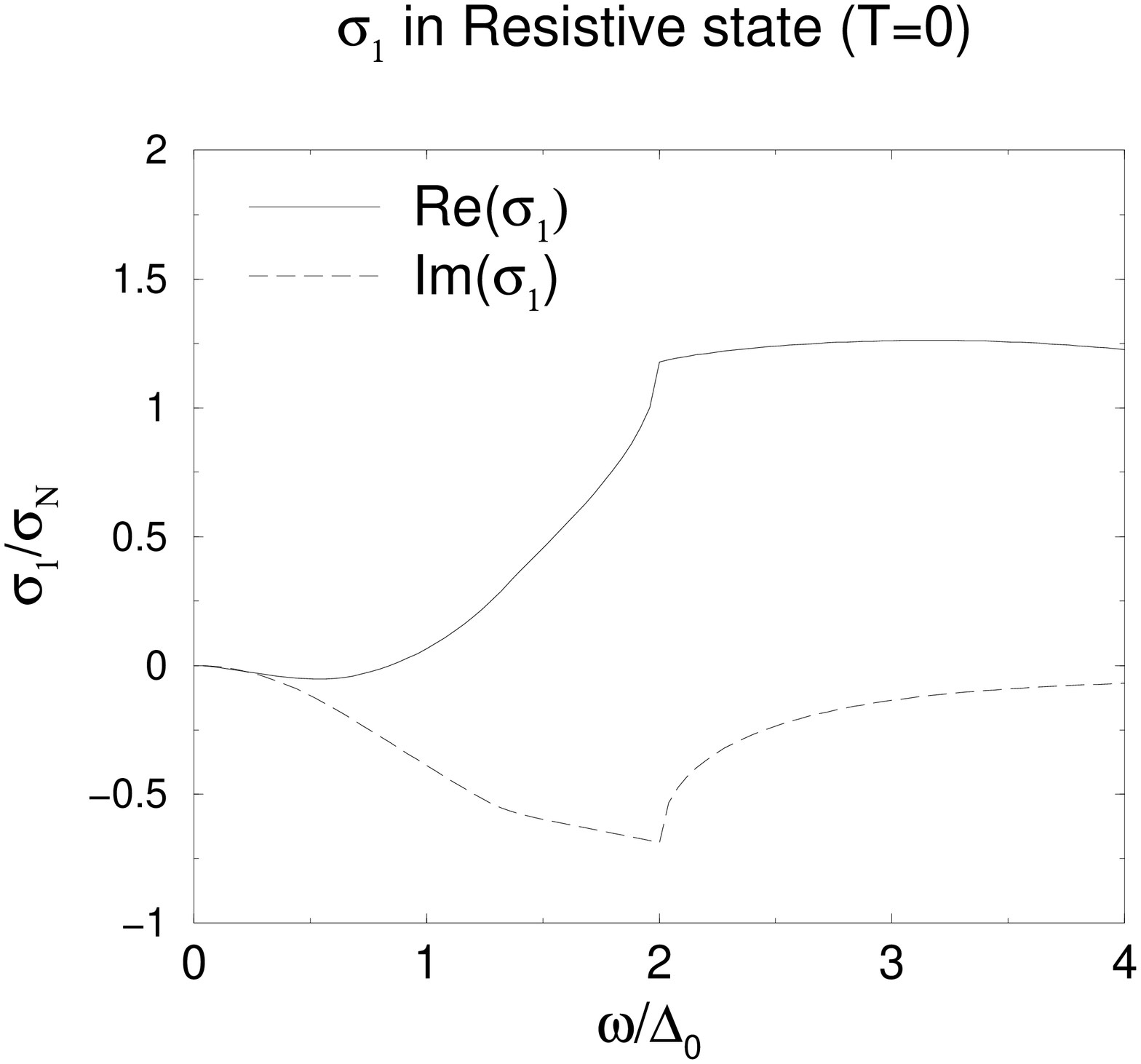,width=0.3\textwidth,angle=270,clip=} 
\end{center} 
\caption{ \label{sigma1} 
Real and imaginary part of $\sigma_1$ in  superconducting 
 ($\Omega=0$, top figure) or resistive  ($\Omega \neq 0$, bottom figure)
junctions  at $T=0$.  
}
\end{figure}

All conductivities $\sigma_i$ are of the same order of magnitude as the 
 experimental  
quasiparticle conductivity $\sigma_{\rm exp} \approx 2 (k 
\Omega {\rm cm})^{-1}$ (Bi-2212) and  
consequently  corrections  $\sigma_i / \epsilon_0 \omega \ll 10^{-2} $ 
are small  
for frequencies $\omega > \omega_{\rm pl} $ on the resistive branch.  
The conductivity $\sigma_1$ is presented for different values of $\Omega$ 
 in fig. \ref{sigma1}. There the transition to the  conductivity 
$\sigma_N = 4 e^2 \pi N_{\rm 2d} t_{s}^2 / \hbar^3$ 
for $\omega > 2 e \Delta_0$ ($\Delta_0$: maximal $d$-wave gap)
and the slight difference between 
the value of $\sigma_1$ in a superconducting and a resistive junction 
can be seen. The negative part for  
small frequencies is not a violation of causality, but arises from the  
fact that $\sigma_1$ is not a conventional transport coefficient.

The static limit of $\sigma_1$ is subtle and deserves special attention.   
Formally the result for $\omega=0$ cannot be obtained in our formalism by  
expanding $\rho_l$ or $j_{l,l+1}$ in the static part
 $\mu_l^{\rm dc}$ of the  
scalar potential directly. Instead of this the static  
inhomogenity has to be included in the equilibrium 
Greens function in zeroth  
order in $t_{\uk \ukp}$. For $\omega \rightarrow 0$ 
and $T < eV \ll \Delta_0$  one gets the finite value
\begin{equation}  
\sigma_1 (\Omega, \omega \rightarrow 0) \sim \sigma_N 
 \frac{T}{\Delta_0}   \frac{eV}{\Delta_0}  \quad ,  
\label{sigma1_limit}
\end{equation}  
while the strictly static function 
$\sigma_1^{\rm dc} := \sigma_1 (\omega =0, T) =0$ vanishes exactly for  
all temperatures.  
This discontinuity is due to the fact that 
our only relaxation mechanism is  the incoherence of the hopping 
$t_{\uk \ukp}$ between layers. Therefore in a tunneling process (which is  
already of order $t_{\uk \ukp}^2$) no additional relaxation within  
the layers is possible, if we work strictly in $O (t_{\uk \ukp}^2 )$. 
Including strong impurity scattering in the layers, which is 
the main mechanism for charge imbalance relaxation in $d$-wave 
superconductors, leads to  $\sigma_1 ( \omega \rightarrow 0)=
\sigma_1 (\omega=0)=0$ like in \cite{artemenko_d}, which will be 
considered mainly in the following. 
It is pointed out that this choice is strictly speaking model dependent
and a finite value $\sigma_{1}^{\rm dc}$ similar to our  result for
$\sigma_1 (\omega \rightarrow 0)$ in equ. \ref{sigma1_limit}  
could be obtained, if additional scattering mechanisms are considered.

The elimination of $V_{\l,l+1}$, $\mu_l$ and $\delta \rho_l$ from equations  
 \ref{poisson}, \ref{josephson}, \ref{strom_allgemein} and  
\ref{ladung_allgemein} 
leads to a set of coupled equations of motion for the phases  
$\gamma_{l,l+1}$ ($\omega_{\rm pl}^2 := 2ed j_c / \hbar \epsilon_0$,  
$\omega_c = \omega_{\rm pl}^2 \epsilon_0 /\sigma_0 $, 
$\Delta_{\alpha}^{(2)} := 1 - \Delta^{(2)}  \alpha $):
\begin{equation} 
\label{equationsofmotion2} \nonumber 
\frac{j}{j_c} = \Delta_{\alpha}^{(2)} \sin \gamma_{l,l+1} +  
\frac{1}{\omega_c}  
    \Delta_\eta^{(2)} \dot{\gamma}_{l,l+1}  + 
   \frac{1}{\omega_{\rm pl}^2}  
   \Delta_{\zeta}^{(2)}   \ddot{\gamma}_{l,l+1}  
 \end{equation} 
The coupling constants are given as: 
\begin{eqnarray} 
  \label{kopplungskonstanten_0} 
  \alpha ( \omega ) &:=&  
    \frac{\epsilon_0 \epsilon}{d \chi_{\rho \rho}} 
  \left( 1 +  \frac{i \sigma_2}{ \epsilon_0 \epsilon  \omega}  \right)  
  \quad , 
  \\ 
  \eta ( \omega ) &:=&  
   \frac{\epsilon_0 \epsilon}{d \chi_{\rho \rho}}  
  \left( 1 +  \frac{i \sigma_2}{ \epsilon_0 \epsilon  \omega} - 
     \frac{2 \sigma_1}{\sigma_0} \right)  \label{eta} 
\quad , 
\\  
 \zeta ( \omega ) &:=& -  \frac{\epsilon_0 \epsilon}{d \chi_{\rho \rho} } 
    \frac{i \sigma_2}{\epsilon_0 \epsilon \omega}  
   \label{kopplungkonstanten3}    . 
\end{eqnarray} 
These are in the static case completely determined by the normal state 
value in equ. \ref{alpha_normal}: 
$\alpha( \omega=0 ) = \eta (\omega =0) = \alpha_0$ and $\zeta (\omega =0) =0$.

The rough theoretical estimate for $\alpha_{0} \approx 0.28$  
(for $\epsilon = 10$, $d = 1.5 {\rm nm}$)  
 based on a twodimensional electron gas   
with density of states $N_{\rm 2d} = m_{\rm el} / 2 \pi^2 \hbar^2$ (not 
including spin) is to be considered as an upper bound, as the density of 
states at the Fermi surface could be enhanced  by a factor 2-3 due to 
strong electronic correlations \cite{dagotto}. A more reliable estimate 
$\alpha \approx 0.2$ is possible from optical experiments and will be 
elaborated in detail  elsewhere \cite{wir_optic}. 


Linearizing the set of equations \ref{equationsofmotion2}  one obtains the 
dispersion of collective modes (in order $O(t_{\uk \ukp}^2)$, $\omega_0 :=  
 \sigma_0 / \epsilon_0 \epsilon = \omega_{\rm pl}^2 /  \omega_c $, 
$\oomega^2 = \omega_{\rm pl}^2 /  \epsilon_c^{\infty}$): 
\begin{eqnarray} 
\label{dispersion_allgemein}  
 \omega^2  
 &\stackrel{k_z \ll 1}{\approx} &  \label{dispersion2} 
\oomega^2 \; \frac{1 + \alpha k_z^2}{1 + \zeta k_z^2} 
      - i \omega \omega_0 \; \frac{1 +  
      \eta k_z^2 }{1 + \zeta k_z^2} = \\ 
      &=&  \oomega^2 \left(  1 + \alpha k_z^2 \right)    
         - i \omega \omega_0 \left(  1 + \eta k_z^2 \right)   
  \quad .  
\end{eqnarray} 
For frequencies $\omega_{\rm pl} \ll \omega \ll \Delta_0 $
on the resistive branch these represent weakly damped plasma
 modes of the superconducting condensate. For the special case 
$e \rightarrow 0$ the  dispersion $\omega (k_z) \rightarrow 0$ for 
$k_z \rightarrow 0 $   
of the Goldstone mode associated with the spontaneous breaking 
of the gauge symmetry in the superconducting state  can be 
obtained explicitly.
This reproduces in the Kuboformalism the results in  
\cite{artemenko_d,artemenko_physicaC}, which were obtained by solving 
kinetic equations.

\begin{figure}[htp!] 
\begin{center} 
\leavevmode 
\epsfig{file=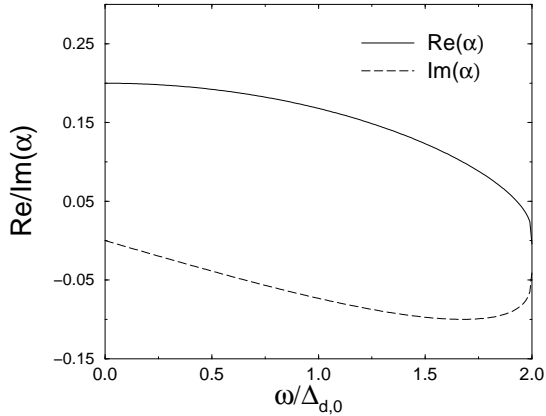,width=0.35\textwidth,angle=270} 
\end{center} 
\caption{ \label{alpha_s} 
Real and imaginary part of $\alpha (\omega)$  
for $d$-wave pairing for temperature $T=0$ in $O(t_{\uk \ukp}^0)$
($\omega$ in units  $\Delta_0 (T)$, $\epsilon_{\infty} = {\rm const.}$).
 } 
\end{figure}

Taking into account only the leading order $O  ( t_{\uk \ukp}^0 ) $ in the  
hopping matrix element, $\alpha (\omega,T) $ has been calculated in 
\cite{diss} and is presented  for $T=0$ in fig. \ref{alpha_s}.  
This result is obtained for a constant background dielectric  
constant $\epsilon$, while the frequency dependence of $\epsilon_{\infty}  
(\omega)$ near phonon resonances can change the behaviour of $\alpha  
(\omega)$ considerably \cite{wir1}. For the temperature dependence see 
\cite{preis_sandiego}.

We would like to point out the remarkable similarity of the {\em static}  
quasiparticle current in the superconducting and in the normal state.  
Due to  $\sigma_{1} ( \omega = 0 )  = 0  $ equ. \ref{strom_allgemein}
 reduces to:  
\begin{equation} 
j_{\rm dc}  =  \nonumber
 j_c \langle \sin \gamma_{l,l+1} \rangle  +
 \frac{\hbar \sigma_0}{2ed}  
\langle {\dot \gamma}_{l,l+1} \rangle  
+ \langle \frac{\sigma_1}{d} (\mu_{l+1}-\mu_l) \rangle 
 \label{jos_static1} 
\end{equation} 
In the last term of equ. \ref{jos_static1} the static part of $\mu$ does not
contribute due to $\sigma_1(\Omega,\omega = 0)=0$, but a 
finite dc-contribution might arise from the combination of $\mu ( \Omega )$ 
and  Josephson-like terms in $\sigma_1$. 
This term is in intrinsic junctions negligible of  order 
$\alpha t_{\uk \ukp}^4 \ll 1$  and vanishes exactly for SN-junctions:  
\begin{eqnarray}
j_{\rm dc}  &=& \left( \langle j_{l,l+1}^{\rm sup} \rangle  +  
\langle j_{l,l+1}^{NN} \rangle  \right) = 
\label{currentstatic}
  \label{currentstatic2}  \\
&=& j_c \langle \sin \gamma_{l,l+1} \rangle  +
\frac{\hbar \sigma_0}{2ed}  
\langle {\dot \gamma}_{l,l+1} \rangle   =  \\ 
&=&  j_c \langle \sin \gamma_{l,l+1} \rangle  +
\sigma_0  \Delta_{\alpha_0}^{(2)} E_{l,l+1}^{\rm dc} 
\end{eqnarray}  
where the quasiparticle current coincides with equ. \ref{currentnormal2}, 
if we express $\gamma_{l,l+1}$ by $E_{l,l+1}$ and $\mu_l$ via 
equ. \ref{josephson}
The dc-equations 
\begin{eqnarray} 
\label{current} 
\nonumber 
j &=&  \sigma_{0}^{SN}  E_{N,N+1}  
+  \left( \sigma_{0}^{SN}-\sigma_{1}^{SN} \right) 
\Delta^{(1)} \mu_N  =  
\label{currentSN_1}
\\ 
&=& \sigma_{0}^{SN} 
 \left( (1+ \alpha_0^{\prime}   ) E_{N,N+1} - \alpha_0^{\prime}  E_{N-1,N}  
 \right) ,  \\  
 j &=& \sigma_0^{SN}
\left( (1+ \alpha_0^{\prime} ) E_{-1,0} - \alpha_0^{\prime}  E_{0,1}  
 \right) \label{currentSN_2}
\end{eqnarray} 
for the SN-junctions are slightly different, as the change of the 
chemical potential $\mu_{-1}^c = \mu^c_{N+1} = 0 $ in bulk materials is 
negligible due to the large density of states. 
Thereby $\alpha_0^{\prime} = \alpha_0 (1 - \sigma_{1,SN}^{rm dc} /
 \sigma_{0,SN}^{\rm dc} )$, 
which is in the model with strong impurity scattering in the layers given 
as: $\alpha_0^{\prime} = \alpha_0$.

If we neglect the supercurrent for a moment, 
the inversion of the  operator $ \Delta_{\alpha_0}^{(2)}$ shows that
the {\em total} voltage $V^{\rm dc} = \sum_{i=-1}^{N} V_{l,l+1}^{\rm dc}$ is 
{\em exactly} given like in a stack of {\em independent} junctions, i.e. 
the coupling constant $\alpha_0$ does not enter in the $I$-$V$-curve.   
The dc-component of the supercurrent might add a correction to this due to
the interaction $\alpha$ at the frequency $\Omega$, but this contribution
is suppressed by  a factor $t_{\uk \ukp}^2 \alpha_0 \ll 1$ and therefore no 
significant effect on dc-properties are expected.

It is pointed out  
that this general  result cannot be obtained by assuming an ohmic  
quasiparticle current $j^{\rm qp}_{l,l+1} = 
\sigma E_{l,l+1}$ as in \cite{tachiki1}. An additional indication that 
the correct choice of the quasiparticle current is essential will come from 
the discussion of Shapiro steps in the next section. 
 
This might also be the reason that up to now no  
clear experimental evidence for the coupling $\alpha$ in the 
$I$-$V$-characteristic,  
like small deviations in the additivity of resistive branches 
 \cite{preis_sandiego},  could be found.  

Nevertheless, the {\em distribution} of the electric fields $E_{l,l+1}^{\rm 
dc} $ within the stack is affected by the coupling $\alpha_0$ like 
in equ. \ref{E_normal1} and \ref{E_normal2}
in the normal state. For a single resistive junction the electric field is 
not contained in this junction, but leaks into neighboring junctions in the 
same way as around a junction with much lower resistance in a stack of 
normal junctions. 


It is stressed that the above discussion makes use of the fact that 
$\sigma_1^{\rm dc} =0$ in the presence of strong impurity scattering in the 
layers. In a more general model this might not be the case and coupling terms
$\sim \alpha_0 \sigma_1^{\rm dc} / \sigma_0^{\rm dc}$ appear.

The coupling via charge fluctuations 
might also have a considerable effect at frequencies larger than the 
charge imbalance relaxation rates (e.g. for phaselocking at microwave
frequencies), where the $\sigma_1 / \sigma_0 $ in equ. \ref{eta} can be of
order $O(1)$ in any model. 

\section{SHAPIRO STEPS}

A Shapiro step is generated by microwave irradiation of frequency $\Omega$,  
which creates an oscillating component of the $c$-axis bias current 
$I_{\rm ac} = I_{\rm rf} \sin (\Omega t)$.  
Thereby the dynamics of the phase in a resistive junction is given  
as  
\begin{equation} 
\gamma_{n,n+1} (t) = \gamma_{0n} + \Omega t  + \gamma_{1n} \sin ( \Omega t  
) 
\label{gammaresistive}  \quad .  
\end{equation} 
Therefore  the supercurrent  
($J_k$: Bessel functions of first kind) 
\begin{eqnarray} 
I_{\rm supra} &=&   I_c \sin \left( \gamma_{n.n+1} (t)   \right) = \\ 
&=&  
I_c  \sum_{k=- \infty}^{\infty} J_k (\gamma_{n1}) \sin \left( \gamma_{n0} + 
(1+k) \Omega t  \right)  
\nonumber
\end{eqnarray}  
has a dc component  
\begin{equation} 
I_{\rm supra}^{\rm dc} = I_c J_{-1} (\gamma_{1n}) \sin ( \gamma_{0n} )   
\quad , 
\end{equation} 
which creates 
 the (first) Shapiro step in the $I$-$V$-curve.  
 

 
The nontrivial question in the following will be,  
what dc-voltage $V_{\rm step}$ is connected with a given $\Omega$,  if  
the generalized Josephson relation equ. \ref{josephson}
is taken into account. Using equation
 \ref{poisson} and \ref{chargegeneral} this 
can be presented as: 
\begin{equation}
\frac{\hbar}{2ed} {\dot \gamma}_{l,l+1} = (1 + 2 \alpha ) E_{l,l+1} - 
    \alpha \left( E_{l-1,l} + E_{l+1,l+2}  \right)
\end{equation}
for $l=0,\dots,N-1$.  
This set of $N$ equations is not sufficient to determine all  
$N+2$ voltages $V_{l,l+1}$ and has to be complemented by the current  
equations \ref{currentSN_1} and \ref{currentSN_2} in the SN-junctions. 

 
Note that for the total voltage $V_{\rm step}$ across the stack  
the transport coefficients in the intrinsic Josephson  
junctions are irrelevant.  
 
In the "superconducting" state, which is defined by $\langle {\dot  
\gamma_{l,l+1}} \rangle = 0$ for all $l=0,\dots,N-1$, one easily obtains 
(in $O(\alpha_0^\prime)$) 
\begin{eqnarray}  
E_{-1,0}^{\rm dc} &=& E_{N,N+1}^{\rm dc} =
 \left( 1-\alpha_0^{\prime} \right)   j / \sigma_{0,SN} \quad ,  \\  
E_{l,l+1}^{\rm dc} &=& 0 \quad \quad l = 1, \dots, N-2 \quad   ,  \\ 
E_{0,1}^{\rm dc} & =&  E_{N-1,N}^{\rm dc} = - \alpha_0^{\prime}  
 j / \sigma_{0,SN}^{\rm dc}  
\end{eqnarray} 
and the total dc-voltage  
\begin{equation} 
\label{contactres}  
V^{\rm dc} = \sum_{l=-1}^N V_{l,l+1}^{\rm dc} = \frac{2 j d}{\sigma_{0,SN}^{\rm dc}}  
\end{equation} 
This linear branch reflects the contact resistances at both SN-junctions  
and is not affected by the coupling $\alpha_0^{\prime}$, although  the
voltages $V_{l,l+1}$ in the intrinsic junctions in general do not vanish.

For a single resistive junction in the top layer $l=N-1$  
($\langle {\dot \gamma}_{N-1,N} \rangle = \Omega$,
$\langle {\dot \gamma}_{n,n+1} \rangle = 0$, $n$ else) 
one obtains analogously:  
\begin{eqnarray} 
\nonumber
E_{N,N+1} &=& (1- \alpha_0^\prime) j / \sigma_{0,SN}^{\rm dc}  + 
\alpha_0^\prime   
\frac{\hbar \Omega}{2ed}  + O(\alpha_2^{\prime 2}) ,  \\  
V_{\rm step}  &=& 
\frac{\hbar}{2e} \Omega \left( 1 - \alpha_0 \frac{\sigma_{1,SN}^{\rm dc}}{ 
\sigma_{0,SN}^{\rm dc}}  \right) \stackrel{\sigma_{1,SN}^{\rm dc}=0}{=} 
\label{vshapzwischen}  
\\  
&=& \frac{\hbar}{2e} \Omega 
\end{eqnarray}  
The last result is correct in all orders of $\alpha_0^{\prime}$.  
Here the dc-conductivity $\sigma^{\rm dc}_{1,SN}=0$ has been kept explicitly
 in equ.  
\ref{vshapzwischen} in order to demonstrate that the position of the 
step in general depends on the value of $\sigma_{1,SN}^{\rm dc}$. 
In the model considered here $\sigma_{1,SN}^{\rm dc}=0$ and the Shapiro step 
is where expected, which would not be obtained by using the ohmic 
quasiparticle current alone ($\sigma_{0,SN}^{\rm dc} = 
\sigma_{1,SN}^{\rm dc}$) like in \cite{tachiki1}.  
Nevertheless this result opens up the principal possibility that there 
{\em is} a shift in the position of the Shapirostep in a more general 
theory (e.g including electron-phonon scattering) due to nonequilibrium 
effects, where $\sigma_1^{\rm dc}$ might not vanish.
Taking out result equ. \ref{sigma1_limit} seriously the relative shift 
would be small $\sim 10^{-5}$, but detectable.

\begin{figure}[htp!] 
\begin{center} 
\leavevmode 
\epsfig{file=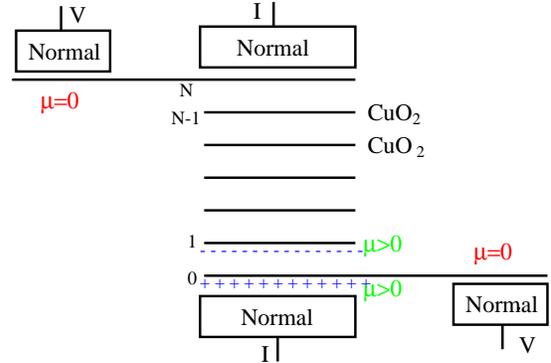,width=0.45\textwidth} 
\end{center} 
\caption{ \label{4point} 
Schematic 4-point geometry  by \cite{latyshev}
} 
\end{figure} 
 
This result is also valid in a  
4-point measurement geometry, as sketched in Fig. \ref{4point} and used  
by \cite{latyshev}. Although the total voltage  
\begin{equation}  
V_{\rm step}^{\rm int} = \sum_{l=0}^{N-1} V_{l,l+1}^{\rm dc} =  
\frac{\hbar \Omega}{2e} \left( 1 - \alpha_0  \right)  + 
\frac{2 j \alpha_0}{\sigma_{0,SN}^{\rm dc} }
\label{4point_v}  
\end{equation}  
across the intrinsic contacts depends on $\alpha_0$, this deviation of the  
expected position of the Shapiro step {\em cannot} be detected at the  
contacts. The electrochemical potential is constant along the  
superconducting layers, but the potential $\mu_l (x) $  
and the electric field $E_{\parallel}$ are not and  will compensate the 
additional contribution in equ. \ref{4point_v}.   
Also in the non equilibrium theory of \cite{ryndyk_jetp} there is no shift 
of Shapiro steps. 
 
Finally, note that all the above considerations only apply to an
experimental situation where the {\em electric field} $E^{\rm dc}$ is 
measured directly to determine the dc-voltage $V^{\rm dc}$. This 
might not be operational with conventional voltmeters, which actually detect 
a current through a circuit with high resistance, which is driven by the 
difference of the electrochemical potentials $\mu_l^{\rm ec}$ rather than the 
voltage $V^{\rm dc}$ \cite{mccumber}.

\section{CONCLUSIONS} 
 
The microscopic theory of tunneling in layered superconductors has been  
studied including the effect of the scalar potential on the layers and an  
estimate of the coupling constant $\alpha \approx 0.2$ 
has been given from optical experiments. The  static
quasiparticle current includes both field and diffusion terms as in the 
normal state, which turns out to be crucial for all dc pro\-per\-ties. 
In the model with strong impurity scattering in the layers,
Shapiro steps are exactly at the expected position 
$V_{\rm step} = \hbar \Omega / 2e$. In this case  the  effect  of charging 
on the total  $I$-$V$-curve  is suppressed by a 
factor  $\alpha_0 t_{\uk \ukp}^2$ and therefore negligible, but it affects 
the distribution of the electric field within the stack. 

This results motivate the study of more general microscopic models including 
different relaxation me\-cha\-nisms like electron-phonon scattering,
as a finite $\sigma_{1}^{\rm dc}$ would modify the position $V_{\rm step}$ 
in a characteristic way with important consequences for  potential 
applications as a voltage standard. Also precision experiments on Shapiro 
steps could then be used as a sharp test for microscopic theories.

The authors would like to thank  S. Rother, R. Kleiner and P. M{\"u}ller  
for discussing their unpublished experimental data and L.N. Bulaevskii  
for his  
interest in this work. Financial support by DFG, FORSUPRA and DOE 
under contract W-7405-ENG-36 
(C.H.) is gratefully acknowledged.

\end{document}